\documentclass[conference]{IEEEconf}

\input epsf
\usepackage{graphicx}
\usepackage{multirow}
\usepackage{titlesec}
\usepackage{balance}
\usepackage{stfloats}
\usepackage{xcolor}

\usepackage{amsmath}
\usepackage{amssymb}
\usepackage{booktabs}
\usepackage{array}
\usepackage{tabularx}
\usepackage{url}
\usepackage{enumitem}
\usepackage{tikz}
\usepackage[hidelinks]{hyperref}
\usetikzlibrary{shapes.geometric, arrows.meta, positioning, calc, fit, backgrounds}

\newcolumntype{Y}{>{\raggedright\arraybackslash}X}
\newcolumntype{C}[1]{>{\centering\arraybackslash}m{#1}}
\newcolumntype{L}[1]{>{\raggedright\arraybackslash\hspace{0pt}}m{#1}}
\newcommand{\thead}[1]{\begingroup\centering\bfseries\shortstack[c]{#1}\par\endgroup}
\providecommand\thesubsectiondis{}
\providecommand\thesubsubsectiondis{}
\providecommand\IEEEoverridecommandlockouts{}
\providecommand\IEEEpeerreviewmaketitle{}
\makeatletter
\@ifundefined{keywords}{%
}{}
\makeatother

\renewcommand\thesection{\arabic{section}}
\renewcommand\thesubsection{\thesection.\arabic{subsection}} % arabic numerals for the subsections
\renewcommand\thesubsubsection{\thesubsection.\arabic{subsubsection}}

\AtBeginDocument{%
  \renewcommand\thesubsectiondis{\thesection.\arabic{subsection}}%
  \renewcommand\thesubsubsectiondis{\thesubsectiondis.\arabic{subsubsection}}%
}
\renewcommand\thesubsectiondis{\thesection.\arabic{subsection}}%

\renewcommand\thesubsubsectiondis{\thesubsectiondis.\arabic{subsubsection}}% arabic numerals for the subsubsections

\titlespacing*{\section}{0pt}{1.0ex}{0.4ex}
\titlespacing*{\subsection}{0pt}{0.8ex}{0.3ex}
\titlespacing*{\subsubsection}{0pt}{0.7ex}{0.25ex}

\makeatletter
\let\IEEEorig@makecaption\@makecaption
\long\def\@makecaption#1#2{%
  \ifx\@captype\@IEEEtablestring
    \footnotesize\bgroup\par\centering\@IEEEtabletopskipstrut
    \parbox{0.98\linewidth}{\centering\normalfont\footnotesize #1. #2}\par
    \egroup\@IEEEtablecaptionsepspace
  \else
    \IEEEorig@makecaption{#1}{#2}%
  \fi
}
\makeatother

\begin{document}

\title{\textbf{\Large JTA: Joint Testability Architecture for Scenario-Based Validation of Safety-Critical Software}}

\author{Wenyao Xue$^{1}$, Jiandi Wang$^{1}$, and Yichen Wang$^{1,*}$\\
\normalsize $^{1}$Beihang University, Beijing, China\\
\normalsize xuewenyao@buaa.edu.cn, 22374306@buaa.edu.cn, wangyichen@buaa.edu.cn\\
\normalsize *corresponding author}

\maketitle

\begin{abstract}
Validation adequacy in safety-critical software depends on more than the system under test. Critical scenarios must be constructed under controlled conditions, execution evidence must be aligned into verdict-ready form, and abnormal outcomes must be attributable to actionable causes. Existing testability research remains largely artifact-centric and offers little architectural support for reasoning about the combined capability of the scenario, the test system, and the system under test. Joint Testability Architecture (JTA) addresses this gap by treating those three elements as a single object of analysis and design. It characterizes validation capability along three dimensions---controllability, observability, and isolability---and organizes them through three domains, three bridges, and an analysis-design-evaluation-refinement loop. JTA also introduces scenario contracts, joint capability assessment, validation blind-spot identification, and bridge-oriented design actions that map capability gaps to concrete improvements in control points, evidence organization, and attribution boundaries. An illustrative analysis of ArduPilot failsafe validation shows that link-loss scenarios are comparatively mature, whereas state-estimation anomaly scenarios remain harder to validate because evidence alignment and attribution semantics are weaker. JTA is not a replacement for existing testing or safety-analysis techniques; it provides an architectural basis for modeling, designing, and assessing scenario-based validation in safety-critical software.
\end{abstract}
\IEEEoverridecommandlockouts
\vspace{1.5ex}
\begin{IEEEkeywords}
Joint Testability Architecture, safety-critical software, test system, controllability, observability, isolability
\end{IEEEkeywords}

\IEEEpeerreviewmaketitle

% ============================
\section{Introduction}
% ============================

In safety-critical systems such as flight control, automated driving, rail transit, and industrial control, validation problems rarely arise from a simple lack of test cases. More often, the difficulty lies in constructing critical scenarios stably, collecting enough execution evidence to support a verdict, or attributing abnormal outcomes to an actionable engineering boundary. This suggests that validation capability is not solely a property of the code under test; it is a system-level capability shaped jointly by the scenario, the test system, and the system under test (SUT).

Traditional research on software testability has accumulated substantial knowledge on controllability, observability, structural complexity, and design-for-testability \cite{voas1995,garousi2019,zakeri2024,iso25010}. However, such research typically interprets testability as a static or semi-static property of the software artifact itself. When the target of analysis shifts to safety-critical scenarios such as failsafe activation, abnormal recovery, mode migration, or multi-component coordinated control, an artifact-centric view is no longer sufficient to answer the following questions: Can the test system reliably construct the target scenario? Can system outputs be organized into verdict-ready evidence? Can validation failures be narrowed down to engineering-actionable responsibility boundaries? In other words, ``code being testable'' does not imply that ``scenario-based validation is adequate.'' 

This gap becomes particularly visible in mature open-source projects. ArduPilot, for example, already provides a relatively complete ecosystem of software-in-the-loop (SITL) simulation, regression testing, and logging infrastructure, together with long-evolved flight-control and failsafe logic. Yet once the validation target shifts from ordinary functional behavior to scenarios such as remote-control (RC) link loss, ground-control-station (GCS) heartbeat timeout, and extended Kalman filter (EKF) state-estimation anomalies, the central question is no longer whether a script exists. It is whether existing scripts, simulation environments, and exposed system hooks can jointly support an adequate validation system.

We therefore introduce Joint Testability Architecture (JTA), which treats testability in safety-critical software as a scenario-conditioned capability of a joint system. The paper addresses three research questions:

\begin{enumerate}[leftmargin=1.5em]
\item \textbf{RQ1:} How can scenarios, the test system, and the system under test be unified into a joint object that is amenable to analysis and design?
\item \textbf{RQ2:} How can controllability, observability, and isolability be operationalized into a repeatable adequacy judgment for scenario-based validation?
\item \textbf{RQ3:} How can joint capability gaps be translated into executable architectural design actions, so that validation blind spots can be systematically reduced?
\end{enumerate}

Our contributions are as follows:

\begin{enumerate}[leftmargin=1.5em]
\item We introduce the joint testability system model $JTS_\sigma=\langle\sigma,T_\sigma,S_\sigma\rangle$, which expands the object of validation in safety-critical software from an isolated software artifact to a scenario-instantiated joint entity composed of the scenario, the test system, and the system under test.
\item We define a scenario contract $SC(\sigma)$ with an explicit fault-source set $r$, thereby turning isolability from a generic slogan into a concrete design constraint centered on distinguishable fault sources.
\item We formulate joint capability and scenario adequacy, and propose a bridge-oriented design method organized as ``three domains, three bridges, and a closed loop,'' so that capability gaps can be mapped directly to concrete design actions on the control bridge, evidence bridge, and attribution bridge.
\item We present an illustrative analysis of ArduPilot failsafe validation, showing how JTA identifies capability blind spots in link-failure and state-estimation anomaly scenarios and derives corresponding bridge-level design recommendations.
\end{enumerate}

The remainder is organized as follows. Section 2 discusses the background and related work. Section 3 presents the Joint Testability Architecture and its formal model. Section 4 describes the capability-gap-driven analysis and design method. Section 5 provides an illustrative evaluation using ArduPilot. Section 6 discusses the novelty, implications, and limitations of the work. Section 7 offers concluding remarks.

% ============================
\section{Background and Related Work}
% ============================

\subsection{Artifact-Centric Testability Research}

Research on software testability has long focused on whether a program is easy to test. Early studies examined how program behavior can be stimulated and observed through controllability and observability \cite{voas1995}. ISO/IEC 25010 includes testability in the software quality model and emphasizes a product's ability to support the establishment of test criteria and the execution of tests \cite{iso25010}. More recent work extends the discussion to design-level testability measurement and architectural transformation for testability improvement \cite{garousi2019,zakeri2024,bagheri2025}. Model-based systems engineering (MBSE) work on design-for-testability (DFT) similarly shows that testability can be shaped during system design \cite{ramirez2024}. Two points from this literature matter most here: controllability and observability remain central dimensions of testability, and testability can be improved through design intervention.

That literature also has a clear scope. The object of analysis is usually the software artifact itself, and the test system and scenario constraints are seldom treated as co-equal analysis objects alongside the system under test. Once the focus shifts from ``whether a piece of code is easy to test'' to ``whether a safety-critical scenario is validated adequately,'' the test system's driving capability, evidence organization capability, and attribution capability become indispensable variables.

\subsection{Safety-Critical Verification and Fault-Driven Assessment}

In safety-critical systems, fault injection, scenario analysis, system safety analysis, and assurance-based validation have all been studied extensively. Fault-injection research uses controlled perturbations to validate fault tolerance and protection mechanisms \cite{arlat1990,hsueh1997}. System safety approaches such as System-Theoretic Process Analysis (STPA), Failure Mode and Effects Analysis (FMEA), and Fault Tree Analysis (FTA) focus on hazards, failure modes, unsafe control actions, and causal scenarios \cite{leveson2012,stpa2018}. More recent work on model-based verification and assurance-case engineering emphasizes organizing evidence and validation activities early in the lifecycle \cite{cederbladh2024,wei2024}. Studies of safety-critical test assessment have also examined test assets through safety tactics and mutation-based fault sensitivity \cite{gurbuz2024}. IEEE Std 1012 frames verification and validation activities from a lifecycle perspective \cite{ieee1012}. Across these lines of work, one point is consistent: validation in safety-critical systems is organized around scenarios, evidence, and accountability chains, not just execution.

JTA is complementary to this literature rather than a replacement for it. Methods such as STPA are effective for identifying critical scenarios and safety constraints, but they do not directly answer whether an existing test system and system under test can jointly support adequate validation for those scenarios. Fault-injection techniques are useful for constructing abnormalities, but they do not by themselves guarantee evidence structuring or attribution convergence. JTA addresses this gap by providing an architectural perspective on \textit{joint adequacy} across these adjacent methods.

\subsection{Test Architecture, Model-Based Testing, and UTP}

Model-Based Testing (MBT) and the UML Testing Profile (UTP) provide standardized ways to represent test contexts, test components, test behavior, and verdict relations \cite{bertolino2003,utp2}. ISO/IEC/IEEE 29119 offers a general framework for testing processes and documentation \cite{iso29119}. Contract-based design validation further shows that test conditions can be expressed explicitly through preconditions and postconditions \cite{meyer1992}. Together, these efforts move the test system beyond an informal collection of scripts and toward an object that can be modeled explicitly. Our recent work on modeling testing requirements for safety-critical software likewise indicates that scenarios, constraints, and testing requirements can be organized systematically \cite{xue2024qrs}.

JTA differs from these approaches in what it takes as the main problem. MBT and UTP focus on how to represent test artifacts and generate testing activities, whereas JTA focuses on how to identify joint capability gaps around critical scenarios and map those gaps to architectural design actions. UTP can serve as one possible representation path for JTA, but it is not the methodological core. The main concern in JTA is \textit{joint adequacy analysis}, not the completeness of modeling notation.

\subsection{Research Positioning}

Existing work provides useful foundations in artifact-level testability, safety-scenario identification, and test modeling. However, there is still no approach that simultaneously satisfies the following three conditions:

\begin{enumerate}[leftmargin=1.5em]
\item it takes the joint entity composed of the scenario, the test system, and the system under test as the object of analysis;
\item it operationalizes controllability, observability, and isolability into an adequacy judgment for scenario-based validation; and
\item it maps validation blind spots directly to executable architectural design actions.
\end{enumerate}

JTA is proposed as a unified response to these three requirements.

% ============================
\section{Architecture and Formal Model}
% ============================

\subsection{Architecture Definition}

The Joint Testability Architecture (JTA) is formalized as
\begin{equation}
JTA = \langle D_{\Sigma}, D_T, D_S, B_C, B_O, B_I, \mathcal{L} \rangle
\end{equation}
where $D_{\Sigma}$ denotes the Scenario Domain, $D_T$ the Test-System Domain, and $D_S$ the System-Under-Test Domain; $B_C$, $B_O$, and $B_I$ denote the control bridge, evidence bridge, and attribution bridge, respectively; and $\mathcal{L}$ denotes the closed-loop process of analysis, design, evaluation, and refinement. Instead of focusing on an isolated software artifact or a standalone testing tool, JTA takes a scenario-centered joint testability system as its primary object of analysis. Plainly, the architecture asks three practical questions for each critical scenario: what must be driven, what must be observed, and what must be distinguished.

\subsection{Joint Testability System}

For project-level validation problems, the joint testability system is defined as
\begin{equation}
JTS=\langle\Sigma,T,S\rangle
\end{equation}
where $\Sigma$ is the set of critical scenarios, $T$ is the test system, and $S$ is the system under test. For any scenario $\sigma\in\Sigma$, the scenario-instantiated joint testability system is defined as
\begin{equation}
JTS_\sigma=\langle\sigma,T_\sigma,S_\sigma\rangle
\end{equation}
where $T_\sigma\subseteq T$ denotes the subset of testing capabilities that actually participate in stimulation, collection, verdict, and diagnosis for scenario $\sigma$, and $S_\sigma\subseteq S$ denotes the boundary of the system under test that is directly relevant to the scenario. This definition emphasizes that the basic object of joint testability is not an isolated software artifact, but a scenario-centered validation assembly. 

\subsection{Scenario Contracts}

Inspired by contract-based formulations that make preconditions and acceptable outcomes explicit in advance \cite{meyer1992}, each critical scenario is represented by the following scenario contract:
\begin{equation}
SC(\sigma)=\langle g,p,u,e,o,r,\kappa,\Theta\rangle
\end{equation}
where:
\begin{itemize}[leftmargin=1.4em]
\item $g$: validation objective of the scenario;
\item $p$: preconditions and environmental constraints;
\item $u$: stimulation mode and stimulus set;
\item $e$: minimum evidence set;
\item $o$: verdict rules;
\item $r$: set of fault sources to be distinguished;
\item $\kappa$: criticality of the scenario;
\item $\Theta=\langle\theta_C,\theta_O,\theta_I\rangle$: minimum requirements on controllability, observability, and isolability.
\end{itemize}

This eight-tuple specifies the minimum semantic boundary required to support valid scenario-based verification. Specifically, $g$, $p$, and $u$ answer ``what is to be validated,'' ``under which constraints,'' and ``by what means the scenario should be activated.'' The pair $e$ and $o$ defines ``which minimum evidence is required'' and ``under what rules a verdict is produced.'' The set $r$ states which potential fault sources must be distinguishable. The threshold vector $\Theta$ specifies the minimum adequacy requirements on controllability, observability, and isolability for the scenario, while $\kappa$ is used to weight risk importance and improvement priority at the project level. Therefore, the scenario contract serves both as the input for deriving capability requirements and as the prerequisite for computing $JTCap(JTS_\sigma)$, $Adeq(\sigma)$, and $Gap(\sigma)$.

Compared with scenario descriptions that mainly serve as testing specifications or requirement statements \cite{xue2024qrs}, JTA explicitly incorporates the fault-source set $r$ into the scenario contract. Safety-critical validation usually cannot stop at determining whether an abnormality has occurred; it must also clarify which candidate sources the abnormality should be distinguished from, so that the validation conclusion can be translated into actionable evidence organization, verdict rules, and rectification boundaries. Without $r$, isolability remains a generic statement and is difficult to transform into explicit constraints on attribution obligations, evidence design, and bridge-level actions \cite{arlat1990,hsueh1997,stpa2018,wei2024}. This is the step that turns a scenario from a testing description into a design constraint.

\subsection{Operationalizing the Three Capability Dimensions}

Assume that the scenario contract $SC(\sigma)$ induces three requirement sets: the control-obligation set $P_C^{req}(\sigma)$, the evidence-obligation set $P_O^{req}(\sigma)$, and the attribution-obligation set $P_I^{req}(\sigma)$. In practice, these sets are a review checklist derived from the scenario contract. For example, a control obligation may be a parameter, injection point, or harness entry; an evidence obligation may be a log field, state snapshot, or event timestamp; and an attribution obligation may be a cause code, source label, or responsibility boundary. In particular, $P_I^{req}(\sigma)$ is jointly determined by $r$ and $o$: if a verdict rule requires the distinction among several fault sources, the system must provide cause points, boundary points, or semantic markers that support such a distinction.

Let the control points, evidence points, and attribution points actually exposed by the system under test be denoted by $P_C^{exp}(\sigma)$, $P_O^{exp}(\sigma)$, and $P_I^{exp}(\sigma)$, respectively. The system-side capabilities are then defined as
\begin{align}
C_S(\sigma)&=\frac{|P_C^{req}(\sigma)\cap P_C^{exp}(\sigma)|}{|P_C^{req}(\sigma)|} \\
O_S(\sigma)&=\frac{|P_O^{req}(\sigma)\cap P_O^{exp}(\sigma)|}{|P_O^{req}(\sigma)|} \\
I_S(\sigma)&=\frac{|P_I^{req}(\sigma)\cap P_I^{exp}(\sigma)|}{|P_I^{req}(\sigma)|}
\end{align}

where $C_S$ reflects how completely the system under test provides controlled activation points for the scenario, $O_S$ reflects the extent to which the system exposes key evidence required for a verdict, and $I_S$ reflects how explicitly it exposes responsibility boundaries and cause semantics. Observability and isolability are therefore related but not identical: observability asks whether relevant facts can be seen, whereas isolability asks whether those facts can separate candidate fault sources into actionable causes.

For cases where a single evidence item is insufficient, JTA represents isolability through conditional distinguishability. For two candidate fault sources $r_i,r_j\in r$, define
\begin{equation}
D_\sigma(r_i,r_j\mid E,\tau)\in\{0,\tfrac{1}{2},1\}
\end{equation}
where $E$ is the available evidence bundle and $\tau$ is the relevant time window. A value of 1 means that the pair can be distinguished directly; $\tfrac{1}{2}$ means that it is distinguishable only when combined evidence or timing constraints are satisfied; and 0 means that the current joint system cannot distinguish the pair. This treatment captures cases such as mode transitions that require both a cause code and a preceding timing pattern before attribution is valid.

The capability of the test system should not be described merely in terms of tool presence or absence. Instead, it should be characterized through repeatable behavioral indicators. Accordingly, nine basic indicators are defined, each taking values in $[0,1]$:
\begin{itemize}[leftmargin=1.4em]
\item $A$: scenario activation success rate; $R$: repeatability consistency; $T_p$: temporal control precision;
\item $E_c$: evidence completeness; $L_a$: evidence alignment quality; $V_a$: verdict automation rate;
\item $D_s$: ratio of distinguishable fault sources; $B_e$: explicitness of responsibility boundaries; $Q_a$: attribution consistency.
\end{itemize}

The test-system-side capabilities are then defined as
\begin{align}
C_T(\sigma)&=\alpha_1A+\alpha_2R+\alpha_3T_p \\
O_T(\sigma)&=\beta_1E_c+\beta_2L_a+\beta_3V_a \\
I_T(\sigma)&=\gamma_1D_s+\gamma_2B_e+\gamma_3Q_a
\end{align}
where $\alpha_i,\beta_i,\gamma_i\in[0,1]$ and each weight set sums to 1. The significance of these equations is that test-system capability depends not only on whether a particular tool exists, but on whether available tools can reliably activate scenarios, produce verdict-ready evidence, and support consistent attribution across executions. Table \ref{tab:operational} summarizes the operational meaning of the three capability dimensions.

\begin{table}[t]
\centering
\caption{Operational Meaning of the Three Capability Dimensions}
\label{tab:operational}
\footnotesize
\setlength{\tabcolsep}{2.5pt}
\renewcommand{\arraystretch}{1.08}
\begin{tabularx}{\columnwidth}{C{2.2cm}YY}
\toprule
\thead{Dimension} & \thead{SUT-side concern} & \thead{Test-system concern} \\
\midrule
Controllability\newline $C$ & Whether sufficient control points are exposed to support controlled scenario activation & Whether the test system can stably construct the scenario, repeat executions, and finely control critical timing \\
Observability\newline $O$ & Whether key events, states, and logs required for a verdict are emitted & Whether the test system can collect, align, structure, and automatically adjudicate evidence \\
Isolability\newline $I$ & Whether explicit responsibility boundaries and cause semantics are exposed & Whether the test system can distinguish fault sources and produce consistent attribution outcomes \\
\bottomrule
\end{tabularx}
\end{table}

\subsection{Scoring and Review Protocol}

To make $JTCap(JTS_\sigma)$ reproducible, the review follows a trace-matrix protocol. Required points are extracted from requirements, scenario contracts, STPA/FMEA/FTA results, verdict rules, and safety constraints; exposed points are extracted from SUT interfaces, parameters, harnesses, log schemas, event streams, cause codes, and test-system adapters. The system-side coverage ratios are then mapped to the ordinal scale used in the case study: H means that at least 80\% of required points are exposed and no critical point is missing; M means 50--80\% coverage or a manual/partly ambiguous evidence chain; L means below 50\% coverage or a missing critical point. For the test-system indicators, equal weights are the default. Non-uniform weights are allowed only when justified by scenario criticality or a safety artifact, and they should be reported with a sensitivity check. This protocol does not remove expert judgment, but it makes the judgment auditable.

\subsection{Joint Capability, Scenario Adequacy, and Validation Blind Spots}

Because scenario-based validation capability is constrained jointly by the test system and the system under test, the joint capability is defined as
\begin{equation}
JTCap(JTS_\sigma)=\langle JC(\sigma),JO(\sigma),JI(\sigma)\rangle
\end{equation}
and
\begin{equation}
\begin{aligned}
JC(\sigma)&=\min(C_S(\sigma),C_T(\sigma))\\
JO(\sigma)&=\min(O_S(\sigma),O_T(\sigma))\\
JI(\sigma)&=\min(I_S(\sigma),I_T(\sigma))
\end{aligned}
\end{equation}

The minimum operator, rather than an average, is used because joint capability is subject to a strong bottleneck constraint. If the system under test does not expose the necessary control points, even a powerful test infrastructure cannot reliably construct the scenario. Conversely, if the test system cannot organize evidence and produce a verdict, the raw logs emitted by the system under test still cannot be converted into a validation conclusion.

Scenario adequacy is defined as
\begin{equation}
Adeq(\sigma)=
\begin{cases}
1, & JC(\sigma)\geq\theta_C \land JO(\sigma)\geq\theta_O \land JI(\sigma)\geq\theta_I \\
0, & \text{otherwise}
\end{cases}
\end{equation}

The capability-gap vector is defined as
\begin{equation}
Gap(\sigma)=\max(\mathbf{0},\Theta(\sigma)-JTCap(JTS_\sigma))
\end{equation}

where the three components of $Gap$ correspond to controllability, observability, and isolability gaps, respectively. The set of validation blind spots is further defined as
\begin{equation}
Blind(\Sigma)=\{\sigma\in\Sigma\mid Adeq(\sigma)=0\}
\end{equation}

To summarize the validation status of critical scenarios at the project level, the system-level joint testability is defined as
\begin{equation}
JT(JTS)=\frac{\sum_{\sigma\in\Sigma}\kappa(\sigma)\cdot Adeq(\sigma)}{\sum_{\sigma\in\Sigma}\kappa(\sigma)}
\end{equation}

where $JT(JTS)\in[0,1]$ denotes the criticality-weighted adequacy coverage over the set of critical scenarios. These definitions are not meant to collapse scenario analysis into a single abstract score. Their immediate role is to answer a more practical question: for a concrete scenario-based validation task, does the current joint testability system meet the minimum adequacy requirements? At the project level, $JT(JTS)$ then provides a traceable summary view.

\subsection{Three Domains, Three Bridges, and a Closed Loop}

JTA organizes the joint testability system around three domains, three categories of bridges, and a continuously evolving closed loop. The ``three'' structure is intentional rather than decorative. The three capability dimensions are necessary because a valid scenario result must be constructed, observed, and attributed; omitting any one of them leaves a different type of blind spot. The three domains are necessary because scenario intent, test-system capability, and SUT exposure are owned by different engineering artifacts. The three bridges are necessary because each cross-boundary mismatch requires a different design response.
\begin{itemize}[leftmargin=1.4em]
\item The \textbf{Scenario Domain $D_\Sigma$} organizes scenario contracts, validation objectives, evidence requirements, verdict rules, and criticality, and serves as the organizing center of the joint design space. 
\item The \textbf{Test-System Domain $D_T$} carries capabilities such as stimulation orchestration, simulation and replay, fault injection, evidence collection, verdict generation, and diagnosis. 
\item The \textbf{System-Under-Test Domain $D_S$} carries not only functional behavior but also the responsibility of supplying validation capability; it therefore needs to expose sufficient control points, evidence points, and responsibility boundaries during design.
\end{itemize}

The three bridges play distinct roles. The control bridge $B_C$ translates the stimulation requirements in the scenario contract into reproducible parameter settings, message injections, temporal control, and constrained anomaly construction. The evidence bridge $B_O$ organizes logs, states, events, and mode transitions into structured evidence and supports automated verdict generation. The attribution bridge $B_I$ maps anomalous outcomes to trigger sources, mode causes, and component boundaries, so that failures can be translated into engineering-actionable remediation tasks. These bridges are architectural patterns, not necessarily domain-independent tools: in concrete systems they may be realized by harnesses, adapters, logging instrumentation, monitors, replay controllers, verdict scripts, or responsibility-boundary annotations.

From the perspective of design operations, the three bridges can be further refined as follows:
\begin{itemize}[leftmargin=1.4em]
\item \textbf{Control bridge $B_C$:} its inputs are the scenario preconditions $p$, the stimulation mode $u$, and the available set of control points; its output is a repeatable sequence of parameterized stimuli. It is activated with priority when $Gap_C(\sigma)>0$. Typical actions include a scenario orchestrator, a timing controller, fault-injection adapters, and injection-enabled replay mechanisms.
\item \textbf{Evidence bridge $B_O$:} its inputs are raw log streams, state streams, and the minimum evidence set $e$; its output is structured, temporally aligned evidence that supports the verdict rules $o$. It is activated with priority when $Gap_O(\sigma)>0$. Typical actions include structured logging, a state snapshot bus, evidence structuring components, and automated verdict rules.
\item \textbf{Attribution bridge $B_I$:} its inputs are failure outcomes, candidate trigger sources, and cause semantics; its output is an attribution result that aligns source, cause, and boundary. It is activated with priority when $Gap_I(\sigma)>0$. Typical actions include cause-code extension, responsibility-boundary mapping, multi-trigger discrimination rules, and semantic annotations for mode transitions.
\end{itemize}

The overall method follows an analysis-design-evaluation-refinement loop: critical scenarios and their capability gaps are first identified; bridge mechanisms are then added according to the gap types; scenario adequacy is subsequently re-evaluated; and any remaining gaps are fed back into the next round of modeling.

\begin{figure*}[t]
\centering
\includegraphics[width=\textwidth]{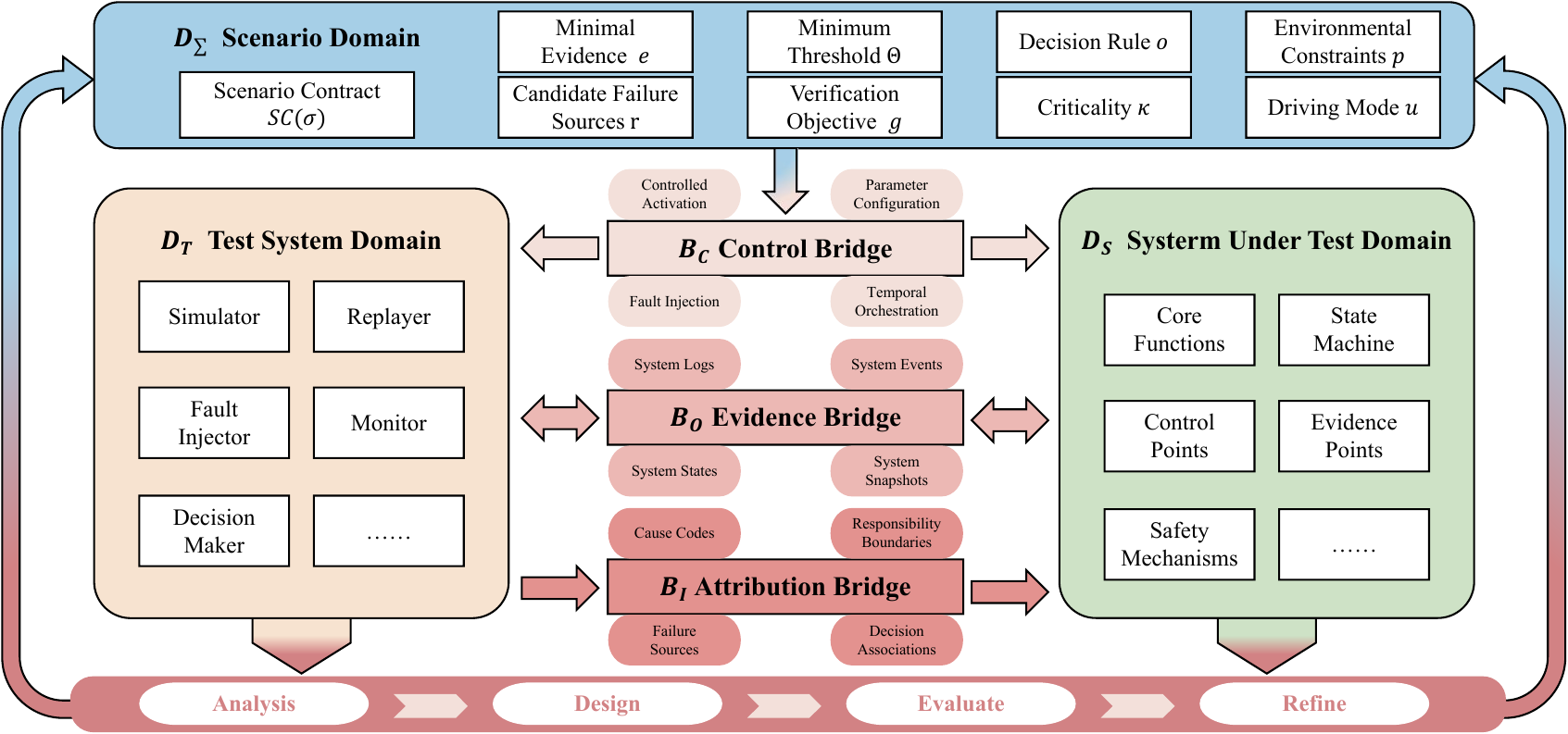}
\caption{The ``three domains, three bridges, and a closed loop'' architecture of JTA. The Scenario Domain organizes contracts and thresholds; the three bridges connect the Test-System Domain and the System-Under-Test Domain; and the closed loop turns capability gaps into evolving bridge-level design actions.}
\label{fig:jta-overview}
\end{figure*}

% ============================
\section{Methodology}
% ============================

\subsection{Five-Step Analysis Workflow}

JTA uses a five-step workflow to analyze critical scenarios. Steps 1--2 instantiate the joint object in RQ1, Step 3 operationalizes the adequacy judgment in RQ2, and Steps 4--5 translate capability gaps into bridge-oriented design actions for RQ3.

\textbf{Step 1: Identify critical scenarios.} Based on requirements, failure reports, domain experience, or the results of system safety analysis such as STPA, FMEA, or FTA, determine the scenario set $\Sigma$ to be assessed and assign a criticality value $\kappa$ to each scenario.

\textbf{Step 2: Instantiate scenario contracts.} For each $\sigma\in\Sigma$, explicitly fill in the objective, preconditions, stimulation mode, minimum evidence set, verdict rules, fault-source set, and threshold vector in $SC(\sigma)$. The focus of this step is not to write an additional test case, but to make explicit the minimum joint conditions required for a valid verification conclusion.

\textbf{Step 3: Assess bilateral capabilities.} Assess $Cap_S(\sigma)=\langle C_S,O_S,I_S\rangle$ and $Cap_T(\sigma)=\langle C_T,O_T,I_T\rangle$ separately. When data are sufficiently available, numerical indicators may be used. In illustrative analyses, each dimension may also be mapped to a three-level ordinal scale---high (H), medium (M), and low (L)---according to a unified rubric.

Table \ref{tab:cap-criteria} provides a set of unified criteria for illustrative assessment, intended to reduce arbitrariness in the H/M/L mapping.

\begin{table}[t]
\centering
\caption{Illustrative Assessment Criteria for the Three Capability Dimensions (H/M/L)}
\label{tab:cap-criteria}
\footnotesize
\setlength{\tabcolsep}{2.5pt}
\renewcommand{\arraystretch}{1.10}
\begin{tabularx}{\columnwidth}{C{1.9cm}YY}
\toprule
\thead{Dimension} & \thead{SUT side ($Cap_S$)} & \thead{Test-system side ($Cap_T$)} \\
\midrule
Controllability\newline $C$ &
\textbf{H}: all critical fault conditions can be triggered independently with configurable timing.\par
\textbf{M}: some conditions are controllable but timing or interfaces are constrained.\par
\textbf{L}: no practical control entry exists for the target condition. &
\textbf{H}: perturbation strength, timing, and duration can be configured independently and repeated reliably.\par
\textbf{M}: the scenario can be driven, but parameters are coupled or repeatability is moderate.\par
\textbf{L}: the target scenario cannot be constructed reliably. \\
Observability\newline $O$ &
\textbf{H}: key states, events, and mode transitions are all emitted with explicit semantics.\par
\textbf{M}: key information is partially visible but evidence is incomplete.\par
\textbf{L}: effective logs or state exposure are missing for critical behavior. &
\textbf{H}: evidence can be collected, temporally aligned, structured, and adjudicated automatically.\par
\textbf{M}: information can be aggregated but manual alignment is still needed.\par
\textbf{L}: effective evidence-processing capability is missing. \\
Isolability\newline $I$ &
\textbf{H}: cause codes cover major trigger paths and responsibility boundaries are explicit.\par
\textbf{M}: partial cause semantics exist but boundaries remain ambiguous.\par
\textbf{L}: cause codes are missing or major sources cannot be distinguished. &
\textbf{H}: consistent source-cause-boundary attribution can be produced.\par
\textbf{M}: major sources can be distinguished but stability is insufficient.\par
\textbf{L}: failure sources cannot be effectively distinguished. \\
\bottomrule
\end{tabularx}
\end{table}

\textbf{Step 4: Compute capability gaps and identify blind spots.} According to the definitions in Section 3, compute joint capability, scenario adequacy, and capability gaps. Any scenario with $Gap(\sigma)\neq\mathbf{0}$ is treated as a validation blind spot and must enter the design stage.

\textbf{Step 5: Perform bridge design and re-evaluation.} Map the gap components to design actions on the control bridge, evidence bridge, and attribution bridge. After the additions are made, re-evaluate scenario adequacy until the minimum requirements are met for critical scenarios or an explicit residual-risk explanation is provided.

\subsection{Capability Gaps and Bridge Design Patterns}

Table \ref{tab:bridge-pattern} summarizes representative symptoms of the three gap categories and their corresponding design patterns.

\begin{table}[t]
\centering
\caption{Capability Gaps and Bridge Design Patterns}
\label{tab:bridge-pattern}
\footnotesize
\setlength{\tabcolsep}{2.5pt}
\renewcommand{\arraystretch}{1.08}
\begin{tabularx}{\columnwidth}{C{1.9cm}YY}
\toprule
\thead{Gap\\dimension} & \thead{Representative\\symptoms} & \thead{Representative\\design actions} \\
\midrule
Controllability\newline $C$ & The target scenario cannot be reproduced stably; critical timing is uncontrollable; perturbations cannot be repeated & Add a scenario orchestrator, parameterized control interfaces, a timing controller, constrained fault-injection adapters, and injection-enabled replay mechanisms to $B_C$ \\
Observability\newline $O$ & Only fragmented logs are available; events and states are not aligned; verdicts depend on manual inspection & Add structured logging, a state snapshot bus, evidence structuring, timeline alignment, and automated verdict rules to $B_O$ \\
Isolability\newline $I$ & Failure sources are unclear; multi-component boundaries are ambiguous; conclusions cannot be turned into remediation tasks & Add cause codes, source labels, responsibility-boundary mapping, mode-transition semantics, and trigger-discrimination rules to $B_I$ \\
\bottomrule
\end{tabularx}
\end{table}

\subsection{Two Adoption Modes}

JTA supports two adoption modes. The \textbf{retrofitting mode} targets existing systems and supplements scenario contracts and bridge mechanisms around critical scenarios on top of available test assets. The \textbf{co-construction mode}, in contrast, uses scenario contracts to constrain the design of control points, evidence points, and responsibility boundaries from the early stage of system design, thereby making validation capability an architectural quality attribute on a par with functional capability. The case study reported here adopts the former mode, but the method itself supports both.

% ============================
\section{Case: ArduPilot}
% ============================

\subsection{Case Objectives and Analysis Protocol}

We use the open-source flight-control stack ArduPilot as an illustrative case \cite{ardupilot_repo}. ArduPilot is a useful case for three reasons. First, the project provides publicly accessible code for flight control, failsafe mechanisms, and logging infrastructure. Second, it already includes SITL, automated testing, and replay capabilities, making it a realistic carrier of an ``existing test system.'' Third, the three scenarios of RC link loss, GCS heartbeat timeout, and EKF state-estimation anomalies span different levels of validation difficulty, from external link failures, to communication-timeout failures, to internal state-estimation degradation. This ordering is intentional: it lets the case show how the bottleneck moves from control, to evidence, to attribution as scenario semantics become more complex.

Beyond illustrating how JTA organizes analysis and design, the case reuses the paper's research questions in a concrete setting: whether JTA can instantiate a joint object for each failsafe scenario (RQ1), whether it can judge adequacy through controllability, observability, and isolability (RQ2), and whether the resulting gaps can be mapped to concrete bridge-level actions (RQ3).

To avoid presenting the case as a runtime evaluation without runtime evidence, the following analysis protocol is adopted:
\begin{itemize}[leftmargin=1.4em]
\item \textbf{$A_0$:} based on the existing ArduPilot codebase and test assets, identify the current joint capability in a structured manner;
\item \textbf{$A_1$:} based on the gaps identified in $A_0$, formulate targeted JTA bridge-level design actions and analyze the expected path for reducing blind spots.
\end{itemize}
We do \textbf{not} report quantitative performance gains after bridge actions are carried out in the system. The case study instead reports the capability assessment and the resulting design-oriented improvement paths around critical scenarios. That choice fits the paper's methodological focus and avoids overcommitting to quantitative claims that the available material does not yet support.

The case maps JTA concepts to concrete ArduPilot assets as follows. The Test-System Domain contains SITL, \texttt{sim\_vehicle.py}, \texttt{autotest.py}, MAVLink heartbeat control, replay assets, and log-processing scripts. The System-Under-Test Domain contains failsafe logic, mode management, EKF modules, DataFlash logging, event messages, and mode-reason enumerations. The Scenario Domain contains the three failsafe contracts in Table \ref{tab:contracts}. Thus the case is not an analysis of the whole repository; it is a scenario-scoped joint testability analysis over the assets that participate in failsafe validation.

The case assessment adopts a unified three-level ordinal rubric:
\begin{itemize}[leftmargin=1.4em]
\item \textbf{H} indicates that the dimension can stably support the core requirements of the scenario contract;
\item \textbf{M} indicates that the dimension partially satisfies the requirements, but still relies on manual supplementation or contains local uncertainty;
\item \textbf{L} indicates that the dimension cannot stably support the core requirements of the contract.
\end{itemize}
To reduce subjectivity in the illustrative assessment, the bilateral capabilities for all scenarios are judged according to the unified criteria in Table \ref{tab:cap-criteria}. To highlight the logic of minimum adequacy, the minimum threshold for all three scenarios is uniformly set to $\Theta=\langle M,M,M\rangle$.

\subsection{Scenario Contract Instantiation}

Table \ref{tab:contracts} lists the key components of the scenario contracts for the three scenarios. Unlike conventional scenario descriptions, the table gives not only the validation objective, stimulation mode, and minimum evidence, but also the explicit fault-source set $r$ that must be distinguished. In this case, $\kappa_{S1}=1.0$, $\kappa_{S2}=1.0$, and $\kappa_{S3}=1.5$ are used to reflect the higher risk significance and validation difficulty of the EKF state-estimation anomaly scenario.

\begin{table*}[t]
\centering
\caption{Key Components of the Scenario Contracts for the Three Scenarios}
\label{tab:contracts}
\footnotesize
\setlength{\tabcolsep}{3pt}
\renewcommand{\arraystretch}{1.08}
\begin{tabularx}{\textwidth}{C{1.15cm}
  >{\hsize=1.00\hsize\raggedright\arraybackslash\hspace{0pt}}X
  >{\hsize=0.90\hsize\raggedright\arraybackslash\hspace{0pt}}X
  >{\hsize=1.10\hsize\raggedright\arraybackslash\hspace{0pt}}X
  >{\hsize=1.00\hsize\raggedright\arraybackslash\hspace{0pt}}X}
\toprule
\thead{Scenario} & \thead{Validation objective $g$} & \thead{Stimulation mode $u$} & \thead{Minimum evidence set $e$} & \thead{Sources to be distinguished $r$} \\
\midrule
S1 RC & Validate whether the expected protective action (RTL/LAND/SMART\_RTL) is triggered after RC link loss & \texttt{SIM\_RC\_FAIL=1}, coordinated with parameters such as \texttt{FS\_THR\_ENABLE} & mode-switch records, failsafe events, \texttt{ModeReason::}\allowbreak\texttt{RADIO\_FAILSAFE}, and recovery path & genuine RC loss; operator misuse at the ground side; test-script timeout or timing drift \\
S2 GCS & Validate whether the specified protective action is triggered after GCS heartbeat timeout & interruption of MAVLink \texttt{HEARTBEAT}, combined with \texttt{FS\_GCS\_TIMEOUT} settings & timeout events, mode-change logs, \texttt{ModeReason::}\allowbreak\texttt{GCS\_FAILSAFE}, and trigger timeline & genuine GCS heartbeat timeout; communication-stack jitter; abnormal heartbeat control in the test toolchain \\
S3 EKF & Validate whether ALT\_HOLD/LAND is triggered when EKF variance or health degrades beyond the threshold & simulation perturbations combined with parameters such as \texttt{SIM\_GPS1\_ENABLE} and \texttt{FS\_EKF\_THRESH} & EKF variance/health status, failsafe events, \texttt{ModeReason::}\allowbreak\texttt{EKF\_FAILSAFE}, and mode-transition timeline & EKF variance beyond threshold; EKF health-bit failure; concurrent GPS quality degradation; mode transition caused by factors other than EKF \\
\bottomrule
\end{tabularx}
\end{table*}

\subsection{Joint Capability Assessment}

Using the code and test-asset mapping established above, the scenario-relevant artifacts can be grouped as follows. On the test-system side, they include \texttt{sim\_vehicle.py}, \texttt{autotest.py}, SITL, the MAVLink heartbeat channel, replay assets, and log-processing assets. On the system-under-test side, they include failsafe logic, mode management, EKF modules, the logging mechanism, and mode-reason enumerations. Following the three-dimensional capability model in Section 3 and the illustrative criteria in Table \ref{tab:cap-criteria}, the three scenarios are assessed ordinally, and the results are shown in Table \ref{tab:capability}. The rating basis is the presence or absence of scenario-specific required/exposed points: controllable parameters and message channels for $C$, aligned logs and mode events for $O$, and cause codes plus source-boundary mappings for $I$.

\begin{figure}[t]
\centering
\begin{tikzpicture}[
    flowstep/.style={rectangle, draw, rounded corners=3pt, fill={rgb,255:red,247;green,227;blue,208}, text width=3.55cm, minimum height=0.82cm, align=center, font=\small, inner sep=4pt},
    flowdec/.style={diamond, draw, fill={rgb,255:red,236;green,184;blue,180}, aspect=2.0, text width=1.9cm, minimum height=0.95cm, align=center, font=\small, inner sep=2pt},
    flowterm/.style={rectangle, draw=green!45!black, rounded corners=3pt, fill=green!16, text width=2.7cm, minimum height=0.82cm, align=center, font=\small, inner sep=4pt},
    arr/.style={-{Stealth[length=2mm]}, thick},
    lbl/.style={font=\scriptsize}
]
\node[flowstep] (s1) at (0,0) {Step 1: Identify critical scenarios\\set $\Sigma$ and $\kappa$};
\node[flowstep] (s2) at (0,-1.70) {Step 2: Instantiate scenario contracts\\fill in $SC(\sigma)$};
\node[flowstep] (s3) at (0,-3.40) {Step 3: Assess bilateral capabilities\\compute $Cap_S,Cap_T$};
\node[flowstep] (s4) at (0,-4.95) {Step 4: Compute gaps\\obtain $JTCap,Gap,Adeq$};
\node[flowdec]  (d1) at (0,-6.65) {$Gap\neq\mathbf{0}$?};
\node[flowstep] (s5) at (0,-8.20) {Step 5: Bridge design\\augment $B_C/B_O/B_I$};
\node[flowterm] (ok) at (4.15,-6.65) {Critical scenarios\\meet adequacy threshold};
\draw[arr] (s1) -- (s2);
\draw[arr] (s2) -- (s3);
\draw[arr] (s3) -- (s4);
\draw[arr] (s4) -- (d1.north);
\draw[arr] (d1.south) -- node[lbl,right,pos=0.52]{yes} (s5.north);
\draw[arr] (d1.east) -- node[lbl,above,pos=0.45]{no} (ok.west);
\draw[arr] (s5.west) -- ++(-1.15,0) |- (s3.west);
\end{tikzpicture}
\vspace{0.8ex}
\caption{The five-step analysis workflow of JTA. Steps 3--5 form an iterative bridge-design loop around capability gaps.}
\label{fig:workflow}
\end{figure}
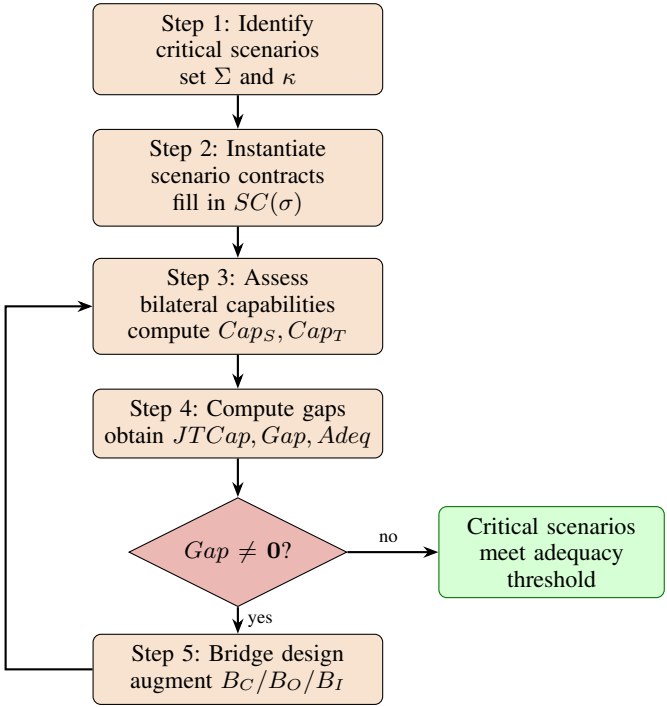

\begin{table*}[t]
\centering
\caption{Joint Capability Assessment Results for the Three ArduPilot Scenarios (H/M/L as Ordinal Ratings)}
\label{tab:capability}
\footnotesize
\setlength{\tabcolsep}{2.8pt}
\renewcommand{\arraystretch}{1.08}
\begin{tabularx}{\textwidth}{C{1.2cm}
  *{9}{C{0.62cm}}
  C{0.78cm}Y}
\toprule
\thead{Scenario} & \thead{$C_S$} & \thead{$O_S$} & \thead{$I_S$} & \thead{$C_T$} & \thead{$O_T$} & \thead{$I_T$} & \thead{$JC$} & \thead{$JO$} & \thead{$JI$} & \thead{$Adeq$} & \thead{Main observations} \\
\midrule
S1 RC & H & H & M & H & M & M & H & M & M & 1 & RC link loss can be triggered reliably. The remaining bottlenecks lie mainly in evidence structuring and coarse-grained cause semantics, but the scenario already meets the minimum adequacy threshold \\
S2 GCS & H & H & M & H & M & L & H & M & L & 0 & Heartbeat timeout can be triggered, but attribution remains weak across timeout occurrence, action trigger, and externally visible cause semantics \\
S3 EKF & M & H & L & L & M & L & L & M & L & 0 & Repeatable construction of EKF anomalies remains weak. Mode transitions are observable, but trigger-source discrimination and attribution consistency are still limited \\
\bottomrule
\end{tabularx}
\end{table*}

Table \ref{tab:capability} yields three observations. First, S1 shows that JTA does not assume every scenario is inadequate. For link-related scenarios with comparatively mature test assets, minimum adequacy can already be achieved. Second, the main problem in S2 is not scenario construction itself, but the weakness of the attribution bridge: the current logs and cause semantics are insufficient to distinguish a genuine GCS heartbeat timeout from control error introduced by the testing toolchain. Third, S3 captures a typical joint-testability challenge. Compared with link failures, state-estimation anomalies depend more heavily on fine-grained perturbation control and cross-component semantic attribution, and therefore reveal bottlenecks in both controllability and isolability.

If the above criticality settings are used to compute system-level joint testability, the current result is
\[
JT(JTS)=\frac{1.0\times 1+1.0\times 0+1.5\times 0}{1.0+1.0+1.5}=28.6\%
\]
This value is a criticality-weighted adequacy coverage computed from the ordinal judgments in Table \ref{tab:capability}; it is not an empirical performance-improvement metric. It shows that the current test assets cover the RC link-loss scenario reasonably well, but substantial blind spots remain in the joint adequacy of the GCS-timeout and EKF-anomaly scenarios. It also shows that unresolved gaps in high-criticality scenarios can significantly lower the project-level validation status, making $JT(JTS)$ a useful indicator for prioritizing improvement efforts.

\subsection{Representative Capability Gaps and Bridge Design Actions}

Building on the above results, Table \ref{tab:gaps} presents representative capability gaps and the corresponding $A_1$ design actions. The table shows how bridge-oriented design turns validation difficulty into explicit engineering action.

\begin{table*}[t]
\centering
\caption{Representative Capability Gaps and Corresponding Bridge-Level Design Actions}
\label{tab:gaps}
\footnotesize
\setlength{\tabcolsep}{3pt}
\renewcommand{\arraystretch}{1.08}
\begin{tabularx}{\textwidth}{C{1.2cm}L{2.0cm}YY}
\toprule
\thead{Scenario} & \thead{Gap\\dimension} & \thead{Gap description} & \thead{Corresponding bridge-level\\design action} \\
\midrule
S2 & Isolability $I$ & The current \texttt{ModeReason::GCS\_FAILSAFE} indicates that a GCS failsafe has occurred, but does not provide an attribution view directly linked to heartbeat counts, timeout thresholds, and link status, which makes it difficult to distinguish genuine timeout from control error in the test environment & Build a triadic attribution view of ``timeout event--cause code--link status'' in the attribution bridge $B_I$; add heartbeat-control logs and timestamp alignment on the test-system side \\
S2 & Observability $O$ & A timing gap remains in the evidence chain between ``heartbeat interruption'' and ``mode transition,'' so verdicts still depend heavily on manual inspection of multi-source logs & Add heartbeat-state snapshots and event alignment in the evidence bridge $B_O$ to structure timeout events, mode transitions, and textual alerts into a coherent timeline \\
S3 & Controllability $C$ & Constructing EKF anomalies depends on simulation perturbations and parameter combinations; perturbation strength, duration, and injection timing are difficult to decouple, leading to insufficient execution consistency & Add parameterized perturbation templates and a timing controller to the control bridge $B_C$ so that perturbation amplitude, duration, and injection timing can be configured independently \\
S3 & Controllability $C$ & Existing replay mechanisms are mainly intended for replaying historical logs and make it difficult to superimpose new perturbations during replay, which limits counterfactual-scenario validation & Add an injection-enabled replay adapter to the control bridge $B_C$ so that real flight data can be combined with controlled perturbations \\
S3 & Isolability $I$ & When ALT\_HOLD/LAND mode transitions occur, the current cause semantics do not clearly distinguish whether the trigger originated from EKF variance beyond threshold, health-status failure, or concurrent GPS quality degradation & Add multi-trigger mapping in the attribution bridge $B_I$ and record EKF variance, health bits, GPS quality scores, and mode reasons jointly as structured cause tuples \\
\bottomrule
\end{tabularx}
\end{table*}

Table \ref{tab:gaps} shows where JTA is most useful. It does not stop at labeling a system ``hard to test'' with a single score; it turns vague engineering impressions into gap items that are dimension-traceable, object-localizable, and action-executable. For S2, the key improvement is not whether the scenario can be triggered one more time, but whether attribution semantics are made explicit. For S3, the key improvement is not whether more logs are added indiscriminately, but whether anomaly construction becomes controllable and cross-component responsibility boundaries become clearer.

\subsection{Findings and Discussion}

The case suggests three points. First, validation difficulty often comes from a mismatch in joint capability: many problems arise from misalignment among scenario requirements, test-system capability, and exposed system hooks, rather than from flight-control code complexity alone. Second, the RC, GCS, and EKF scenarios exhibit different bottleneck structures, which supports treating testability as a scenario-conditioned property rather than a single system-level attribute. Third, existing SITL, AutoTest, and logging infrastructure already provide a substantial starting point; JTA helps reorganize those assets so that they move from merely ``being able to run'' toward ``being able to adjudicate, attribute, and refine.''

% ============================
\section{Discussion}
% ============================

\subsection{Novelty and Research Positioning}

The main novelty lies in three shifts. First, the object of testability analysis moves from an isolated software artifact to the triplet $JTS_\sigma$ composed of the scenario, the test system, and the system under test. Second, a scenario is treated not merely as a descriptive item in a test specification, but as a minimal design unit that carries objectives, evidence requirements, verdict rules, a fault-source set, and minimum thresholds through the scenario contract. Third, architectural action becomes an explicit outcome of analysis: capability gaps are mapped directly to design actions on the control bridge, evidence bridge, and attribution bridge. The novelty is therefore not the isolated idea that fault sources can be traced, which also appears in safety analysis and MBSE practice; it is the use of fault-source obligations inside a scenario-conditioned joint adequacy review that covers the scenario, the test system, and the SUT together.

\subsection{Implications for Safety-Critical Software Engineering}

JTA treats dependability testing as an architectural issue rather than a purely execution-level one. In practical terms, validation capability should not be checked only after functional implementation is complete; it should be organized proactively through scenario contracts during design. For safety-critical software, this shift matters in at least four ways. First, it upgrades the test system from an auxiliary collection of scripts to a formal design object. Second, it turns evidence organization and responsibility boundaries into first-class design concerns. Third, it provides a finer-grained common language for assurance-case engineering, shift-left validation, and cross-team collaboration \cite{cederbladh2024,wei2024}. Fourth, it complements the lifecycle-oriented view of verification and validation required by IEEE Std 1012 \cite{ieee1012}.

\subsection{Limitations and Threats to Validity}

This study also has clear boundaries. First, it is a method-and-architecture study. The purpose of the case is to demonstrate how JTA organizes analysis and design, not to report performance gains or certification benefits after all bridge actions are carried out. Second, the H/M/L ordinal assessment adopted here is suitable for illustrative analysis under the current material conditions, but it still involves engineering judgment. Future work may refine it into automatically extractable numerical indicators and more rigorous review rubrics. Third, the case comes from the open ArduPilot ecosystem, whose openness and testing-infrastructure maturity are project-specific. The applicability of JTA in closed industrial systems, especially under restricted control interfaces, restricted logging, or restricted evidence disclosure, still requires further case-based validation.

The main validity risks of the present study fall into four categories.
\begin{itemize}[leftmargin=1.4em]
\item \textbf{Construct validity:} the current H/M/L ordinal assessment is still an illustrative discretization of the numerical model in Section 3. In future work, indicators such as control-interface completeness, log coverage, and cause-code coverage should be further automated.
\item \textbf{Internal validity:} the assessment process inevitably contains researcher judgment, especially in defining scenario boundaries and minimum evidence sets.
\item \textbf{External validity:} the openness and testing-infrastructure maturity of ArduPilot are higher than those of many industrial projects; the conclusions therefore cannot be generalized directly to all safety-critical software.
\item \textbf{Method boundary:} JTA is intended as a meso-level architectural analysis and design method. It does not replace formal verification, certification processes, or concrete testing frameworks; rather, it provides a unified perspective for organizing validation capability across them.
\end{itemize}

% ============================
\section{Conclusion}
% ============================

Joint Testability Architecture (JTA) reframes testability in safety-critical software as a scenario-conditioned system capability carried jointly by the scenario, the test system, and the system under test. The paper defines the joint testability system, scenario contracts, joint capability, scenario adequacy, validation blind spots, and the system-level joint testability metric $JT(JTS)$, and uses the ``three domains, three bridges, and a closed loop'' formulation to map capability gaps to concrete design actions on the control bridge, evidence bridge, and attribution bridge.

The illustrative analysis of ArduPilot failsafe validation shows that the main bottlenecks of link-related scenarios lie in evidence structuring and the explicitness of cause semantics, whereas the main bottlenecks of EKF anomaly scenarios lie in the controllability of anomaly construction and the consistency of cross-component attribution. This result indicates that ``difficulty of testing'' in safety-critical software is not merely a code-level issue, but is caused by joint mismatches among scenario requirements, testing infrastructure, and exposed system hooks.

The broader implication is that, in safety-critical software, validation capability should be designed, not merely checked after the fact. That means bringing the test system into architectural decisions, treating the scenario as the organizing unit of validation, and evaluating testability as a joint system capability. Future work will refine capability indicators in additional industrial cases, validate the benefits of implementing bridge-level actions, and explore deeper integration with UTP, assurance cases, and model-driven engineering.

\balance

\end{document}